\documentclass[aps,prl,reprint,showpacs,superscriptaddress,nobibnotes,nofootinbib]{revtex4-1}
\usepackage{graphicx}
\usepackage{amssymb} 
\usepackage{amsmath}
\newcommand{\vect}[1]{\boldsymbol{#1}}
\newcommand{\minisec}[1]{{\bf #1}\\}
\begin{document}

\title{Precise Measurement of Magnetic Field Gradients from Free Spin Precession Signals of $^{3}$He and $^{129}$Xe Magnetometers}
\author{F. Allmendinger}
\email{Corresponding author: allmendinger@physi.uni-heidelberg.de}
\affiliation{Physikalisches Institut, Ruprecht-Karls-Universit\"{a}t, Heidelberg, Germany}

\author{P. Bl\"{u}mler}
\affiliation{Institut f\"{u}r Physik, Johannes Gutenberg-Universit\"{a}t, Mainz, Germany}

\author{M. Doll}
\affiliation{Institut f\"{u}r Physik, Johannes Gutenberg-Universit\"{a}t, Mainz, Germany}

\author{O. Grasdijk}
\affiliation{Van Swinderen Institute, University of Groningen, The Netherlands}

\author{W. Heil}
\affiliation{Institut f\"{u}r Physik, Johannes Gutenberg-Universit\"{a}t, Mainz, Germany}

\author{K. Jungmann}
\affiliation{Van Swinderen Institute, University of Groningen, The Netherlands}

\author{S. Karpuk}
\affiliation{Institut f\"{u}r Physik, Johannes Gutenberg-Universit\"{a}t, Mainz, Germany}

\author{H.-J. Krause}
\affiliation{Peter Gr\"unberg Institute, Forschungszentrum J\"ulich, Germany}

\author{A. Offenh\"{a}usser}
\affiliation{Peter Gr\"unberg Institute, Forschungszentrum J\"ulich, Germany}

\author{M. Repetto} 
\affiliation{Institut f\"{u}r Physik, Johannes Gutenberg-Universit\"{a}t, Mainz, Germany}

\author{U. Schmidt}
\affiliation{Physikalisches Institut, Ruprecht-Karls-Universit\"{a}t, Heidelberg, Germany}

\author{Yu. Sobolev} 
\affiliation{Institut f\"{u}r Physik, Johannes Gutenberg-Universit\"{a}t, Mainz, Germany}

\author{K. Tullney}
\affiliation{Institut f\"{u}r Physik, Johannes Gutenberg-Universit\"{a}t, Mainz, Germany}

\author{L. Willmann}
\affiliation{Van Swinderen Institute, University of Groningen, The Netherlands}

\author{S. Zimmer}
\affiliation{Institut f\"{u}r Physik, Johannes Gutenberg-Universit\"{a}t, Mainz, Germany}

\date{\today}
\begin{abstract}
We report on precise measurements of magnetic field gradients extracted from transverse relaxation rates of precessing spin samples. The experimental approach is based on the free precession of gaseous, nuclear spin polarized $^3$He and $^{129}$Xe atoms in a spherical cell inside a magnetic guiding field of about 400 nT using LT$_C$ SQUIDs as low-noise magnetic flux detectors. The transverse relaxation rates of both spin species are simultaneously monitored as magnetic field gradients are varied. For  transverse relaxation times reaching 100 h, the  residual longitudinal field gradient across the spin sample could be deduced to be$|\vect{\nabla}B_z|=(5.6 \pm 0.4)$ pT/cm.
The method takes advantage of the high signal-to-noise ratio with which the decaying spin precession signal can be monitored that finally leads to the exceptional accuracy to determine magnetic field gradients at the sub pT/cm scale. 
\end{abstract}

\maketitle

\minisec{Introduction}
$^3$He magnetometers based on free spin precession provide ultra-sensitive measurements and monitoring of magnetic fields as demonstrated recently in \cite{Gemmel,Koch,Nikiel}.  For the readout of the spin precession signal one can use several sensors like low- or high-T$_C$ SQUID gradiometers, Rb or Cs gradiometers or standard NMR techniques. At low magnetic fields ($B_0< 50~\mu$T) it is advantageous to use SQUIDs or alkali-magnetometers to record the free spin precession since they directly measure the temporal change of the $^3$He magnetization $M(t)$. At magnetic fields exceeding 0.1 T, NMR detection techniques are clearly preferable because they detect the induced field of the precessing sample magnetization being proportional to $dM/dt$, i.e., the recorded signal scales with the Larmor frequency and thus with the magnetic field strength.\\
Optical pumping is the technique used to hyperpolarize diluted noble gases for sufficient signal enhancement resulting in a high Signal-to-Noise Ratio (SNR). Whereas metastability optical pumping (MEOP) \cite{Schearer,Courtade} is used to hyperpolarize the $^3$He nuclear spins along the axis of the respective magnetic field $\vect{B_0}$ (z-axis), the second spin species for our comagnetometry studies,  $^{129}$Xe, is spin-polarized by spin exchange optical pumping (SEOP) \cite{Walker}.\\
 Then the nuclear spins are tipped synchronously out of axis towards the transverse x-y plane by applying a short, resonant radio frequency pulse, or by non-adiabatic spin flipping. Subsequently, the free, coherent precession of the nuclear magnetic moments around the field direction with the Larmor frequency 
\begin{eqnarray}\omega_L&=&\gamma \cdot B_0
\label{eqn:Larmor}
\end{eqnarray}
is detected. The proportionality constant $\gamma$ is called the gyromagnetic ratio and is a property of the respective nucleus ($\gamma_{He} = -2\pi\cdot32.434 099 66(43)$ MHz/T \cite{Mohr}).\\
The presence of a magnetic field gradient in a sample cell containing spin-polarized noble gases will increase the transverse relaxation rate. The origin of this relaxation mechanism is the loss of phase coherence of the atoms due to the fluctuating magnetic field seen by the atoms as they diffuse throughout the cell. In the motionally narrowing regime, where the gas atoms diffuse throughout the entire sample cell (spherical cell of radius $R$) in a relatively short time $\tau_D\approx R^2/D\ll 1/(\gamma \Delta B)$, the perturbing influence of the field inhomogeneity $\Delta B \approx R\cdot |\vect{\nabla B}|$ on the spin coherence time $T_2^*$ is strongly suppressed. Analytical expressions can be derived for the transverse relaxation rate for spherical sample cells, as reported in \cite{Cates1}. Subsuming the relaxation time at the walls, $T_{1,wall}$, and other spin-relaxation modes under the longitudinal relaxation time $T_1$, the general expression for the transverse relaxation rate $1/T_2^*$ for a spherical sample cell of radius $R$ is
\begin{eqnarray}
\nonumber
\frac{1}{T_2^*}=
\frac{1}{T_1}+\frac{8R^4\gamma^2}{175D}\left(|\vect{\nabla}B_z|^2+a(\lambda)\cdot\left(|\vect{\nabla}B_x|^2+|\vect{\nabla}B_y|^2\right)\right)\\
~
\label{eqn:T2*}
\end{eqnarray}
with
\begin{eqnarray}
a(\lambda)&=&\frac{175}{8}\lambda \cdot \sum_n \frac{1}{|x_{0,n}^2-2|\cdot \left(1+x_{0,n}^4 \lambda\right)}
\label{eqn:a}
\end{eqnarray}
and
\begin{eqnarray}
\lambda&=&\frac{D^2}{\gamma^2 B^2 R^4}~~.
\label{eqn:lambda}
\end{eqnarray}
Here, $D$ is the diffusion coefficient of the gas and $x_{0,n}$ ($n = 1, 2, 3, . . .$) are the zeros of the derivative$\frac{d}{dx} j_1(x) = 0$ of the spherical Bessel function $j_1(x)$. The deviation $\vect{B'}(\vect{r})$ of the local field from the average homogeneous field $\vect{B_0}$ was approximated by the uniform gradient field $\vect{B'}(\vect{r})= \vect{G}\cdot \vect{r}$, with $\vect{G}$ being a traceless, symmetric second-rank tensor.\\
Eq.~(\ref{eqn:T2*}) above suggests measurements at low gas pressures ($D\propto 1/p$) and at small sample sizes ($T_2^*\propto R^{-4}$). However lowering both gas pressure and size reduces SNR and thus the measurement sensitivity. As shown in \cite{Gemmel}, optimum conditions are met at gas pressures of a few mbar and sample sizes of several cm. In such a way it is possible to obtain characteristic times of coherent spin precession of up to 100 h in homogeneous magnetic fields below $1~\mu$T using almost relaxation free sample containers with $T_1 > 100$ h. Even at high magnetic fields (above $0.1$ T), $T_2^*$ of several minutes has been measured \cite{Nikiel}.\\
Besides ultra-precise monitoring of magnetic fields ranging from nT $< B_0 < 10$ T, the detection of the free spin precession also provides direct access to  magnetic field gradients via the measurement of the exponential decay of the recorded signal amplitude $S\propto \exp(-t/T_2^*)$. This option has not been systematically followed in the past because the focus was primarily laid on the exploration of precise measurement and monitoring of the magnetic field modulus extracted from the measured Larmor frequency (Eq.~(\ref{eqn:Larmor})). With $T_2^*$ as an additional observable, access to all nine tensor elements is possible, however, only five are independent as a result of Maxwell's equations in free space. 
By inspection of Eq.~(\ref{eqn:T2*}) it is clear that by varying the parameter $\lambda$, e.g., by changing the gas pressure $p$, the relative contribution of the transversal field gradients $|\vect{\nabla}B_x|$ and $|\vect{\nabla}B_y|$ to $T_2^*$ can be adjusted. The weighing can be inferred from Fig.~1, where the pre-factor $a(\lambda)$ from Eq.~(\ref{eqn:a}) is plotted as a function of $\lambda$. \\
\begin{figure}
\includegraphics[scale=1]{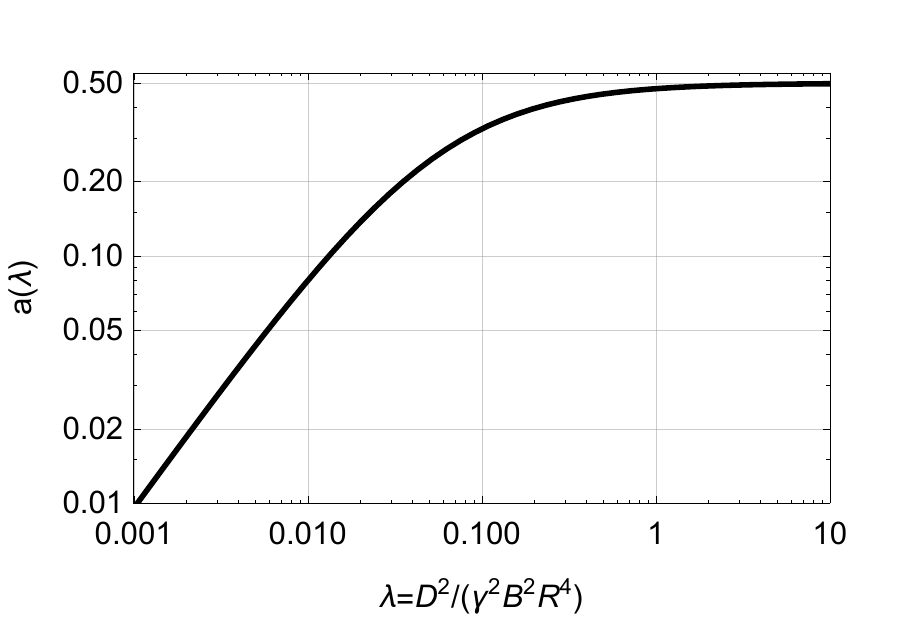}
\caption{The pre-factor $a$ of Eqs.~(\ref{eqn:T2*}) and (\ref{eqn:a}) as a function of $\lambda$. By varying the parameter $\lambda$, e.g., by changing the gas pressure $p$, the relative contribution of the transversal field gradients $|\vect{\nabla}B_x|$ and $|\vect{\nabla}B_y|$ to $T_2^*$ can be adjusted.}
\label{fig:ax}
\end{figure}
Accurate determination of magnetic field gradients, in particular the transverse gradient components, is a challenging metrological task: sensor offsets and their temporal drifts, non-orthogonality of the sensor axes and misalignment angles as result from imperfections of the mechanical design of the mapper often severely limit the measurement accuracy. To motivate the need for an accurate and precise control of field  gradients: At current and anticipated levels of sensitivity in electric dipole moment (EDM) measurements, geometric-phase-induced false EDM signals, resulting from interference between magnetic field gradients and particle motion in electric fields, are an important potential source of systematic errors \cite{Pendlebury}. Precise and accurate gradient measurements of order pT/cm are demanded to correct directly for these false-EDM signals in the future \cite{Afach}.\\
Here, we report on a first field gradient measurement via recorded  $T_2^*$-times using a $^3$He-$^{129}$Xe co-magnetometer. A co-located $^3$He and $^{129}$Xe spin sample is used to verify, in the practical implementation, the validity of the analytical expression for $T_2^*$ given in  Eq.~(\ref{eqn:T2*}). This serves as a solid basis to finally extract precise numbers for field tensor components from $T_2^*$ measurements. With the chosen gas pressures and magnetic field ($B_0\approx 400$ nT) we meet the situation $a(\lambda) \approx 0$. Then the longitudinal field gradient component $|\vect{\nabla}B_z|$ across the sample volume ($V=456~\text{cm}^3$) can be extracted from the measured $T_2^*$-times for different settings of the applied magnetic field.\\ 
\minisec{Experimental setup and procedure}
The experimental setup was described in detail in \cite{Gemmel, Heil, Allmendinger}. Briefly, the basic setup consists of the low-relaxation spherical measurement cell ($R=4.8$ cm) filled with a gas mixture of polarized $^3$He ($p_{He}=3.4$ mbar), polarized Xe ($85\%~^{129}\text{Xe},~p_{Xe}=4.9$ mbar) and N$_2$ ($p_{N_2}=24.5$ mbar) as a buffer gas. The sample cell is brought into a 7-layered magnetically shielded room (BMSR-2, \cite{Bork}) and placed directly below the Dewar housing the LT$_C$- SQUID vector magnetometer system. This system detects a sinusoidal magnetic field change due to the spin precession of the gas atoms. After degaussing, the residual magnetic field of the BMSR-2 is estimated to be about 1 nT. Two square Helmholtz coil pairs arranged perpendicular to each other with adjustable current sources (resolution: 100 nA, noise density: $82~\text{pA}/\sqrt{\text{Hz}}$,  stability $10^{-4}$) generate a magnetic guiding field of $|\vect{B_0}|=403$ nT. The guiding field - and with it the quantization axis $z$ - can be oriented in any direction $\alpha$ in the horizontal plane keeping $B_0$ constant to a level of 1 nT. The coil system also serves to manipulate the sample spins, e.g., generating a $\pi$/2 spin-flip by non-adiabatic switching. The magnetic field gradients are varied by turning the guiding field inside the magentically shielded room. There are two main sources of gradients: Residual field gradients from the mu-metal shielding and gradients produced by the Helmholtz coils. The latter ones will change, as the magnetic guiding field is rotated. The resulting gradients are the sum of the two above. In the chosen experimental procedure, the magnetic guiding field is rotated slowly for a certain amount $\Delta\alpha=45^\circ$ in 5~minutes and then stays constant for 25 minutes. In that phase of operation, the SQUID system detects the field of the precessing magnetization of the polarized gases. By exponential fits to the decaying He and Xe amplitudes, $T_{2,He}^*$ and $T_{2,Xe}^*$ are determined for different values of $\alpha=n\cdot \Delta\alpha$, where we set $\alpha = 0$ for the guiding magnetic field pointing parallel to the entrance door wall.\\
\minisec{Results}
\begin{figure}

\includegraphics[scale=1]{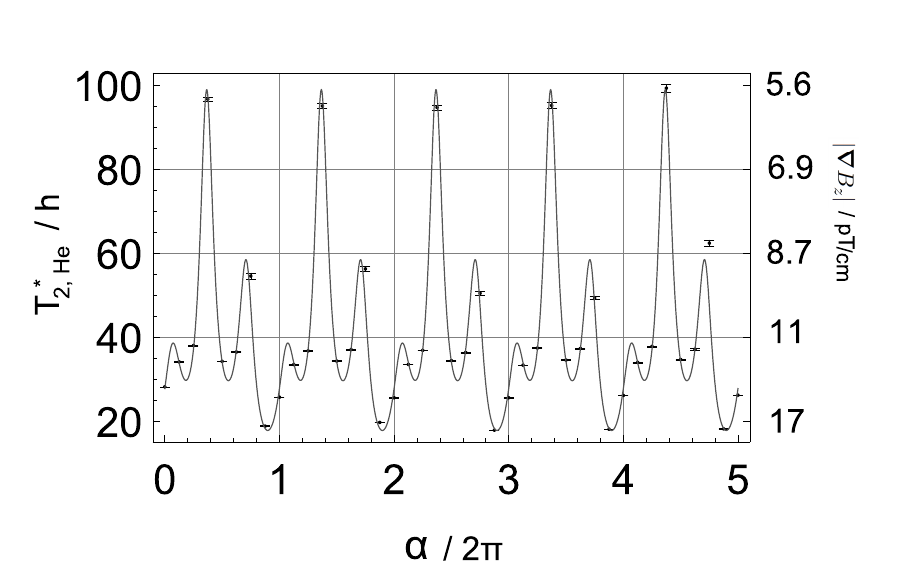}
\includegraphics[scale=1]{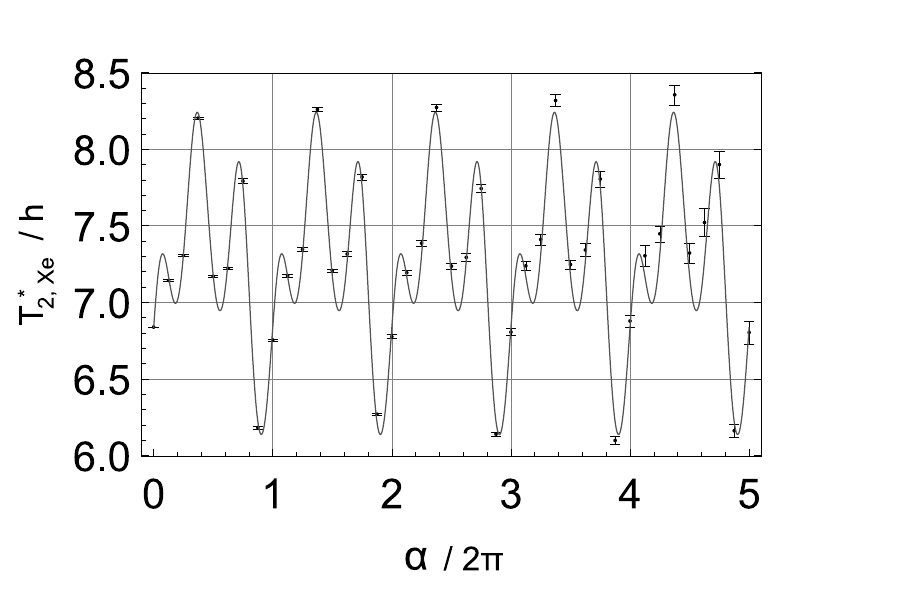}
\caption{The transverse relaxation times of helium and xenon as a function of the direction $\alpha$ of the magnetic guiding field in the horizontal plane, applied in steps of $\Delta\alpha=45^\circ$ for 5 turns. In total, the measurement took  about 20 hours with $\sim$30 min for each field setting to extract $T_2^*$ from the exponential decay of the signal amplitude. Solid line: Fit of a Fourier series to the measured relaxation rates $1/T_2^*$ to guide the eye. From the transverse relaxation times, the respective magnetic field gradients can be extracted according to Eq.~(\ref{eqn:T2*2}). For the given experimental parameters, the corresponding range of the longitudinal field gradient $|\vect{\nabla}B_z|$ is shown on the right hand side of the $T_{2,He}^*$ plot. 
}
\label{fig:T2alpha}
\end{figure}
The transverse relaxation times for $^3$He and $^{129}$Xe are shown in Fig. \ref{fig:T2alpha}. They depend strongly on the direction of the magnetic guiding field $\alpha$ and vary between 20 h and 100 h for Helium, and between 6 h and 8.5 h for Xenon. The characteristic pattern in Fig. \ref{fig:T2alpha} repeats itself after every revolution. At some angle $\alpha$,  the gradients from the chamber and coils almost cancel each other and $T_2^*$ is maximized. At other angles the cancellation is less distinct with a minimum in $T_2^*$ at a field orientation where the gradients add up constructively. This is consistent with the observation that the rotation of the magnetic guiding field by $180^\circ$ changes the transverse relaxation time from the global maximum to the global minimum.\\
The precision in extracting the transverse spin relaxation times is a few percent, and this despite the fact that we only see a relatively weak decay of the signal amplitude $\Delta S/S=-\Delta t/T_2^*\approx -4\cdot 10^{-3}$ (linear term of the exponential decay) during the data acquisition time of $\Delta t=25$ min in case of $T_2^*$ approaching 100 h. The reason for this high detection sensitivity is the excellent SNR of 4000:1 in a bandwidth of 1 Hz, resulting in a precision of amplitude determination of $\delta S/S\approx 7\cdot 10^{-6}$ after this relatively short acquisition time. Drifts of the transverse relaxation time of $\Delta T_2^* / \Delta t> 1~\text{min}~/~30~\text{min}$ caused by temporal changes of the field gradients will result in an increased $\chi^2$-value of the exponential fit to the data. This consequence was not observed, and indeed former measurements using coherent spin precession which monitored the decaying signal amplitude over extended periods of $\Delta t \approx T_2^*$ confirmed this finding by $\Delta T_2^* < 160$~s \cite{Tullney}.\\
In the expressions of Eq.~(\ref{eqn:T2*}), the term including the gradients can be eliminated in case of simultaneous measurements of the relaxation rates $1/T_{2,He}^*$ and $1/T_{2,Xe}^*$ leading to:
\begin{eqnarray}
\frac{1}{T_{2,He}^*}&=&k+m\frac{1}{T_{2,Xe}^*}
\label{eqn:T2*compare}
\end{eqnarray}
with
\begin{eqnarray}
k&=&\frac{1}{T_{1,He}}-m\frac{1}{T_{1,Xe}}
\label{eqn:k}
\end{eqnarray}
and
\begin{eqnarray}
m&=&\frac{\gamma_{He}^2 D_{Xe}}{\gamma_{Xe}^2 D_{He}}~~.
\label{eqn:m}
\end{eqnarray}
Fig. \ref{fig:HevsXe} shows the measured pairs of relaxation rates that clearly follow a straight line with slope $m$ and intercept $k$.
\begin{figure}
\includegraphics[scale=1.]{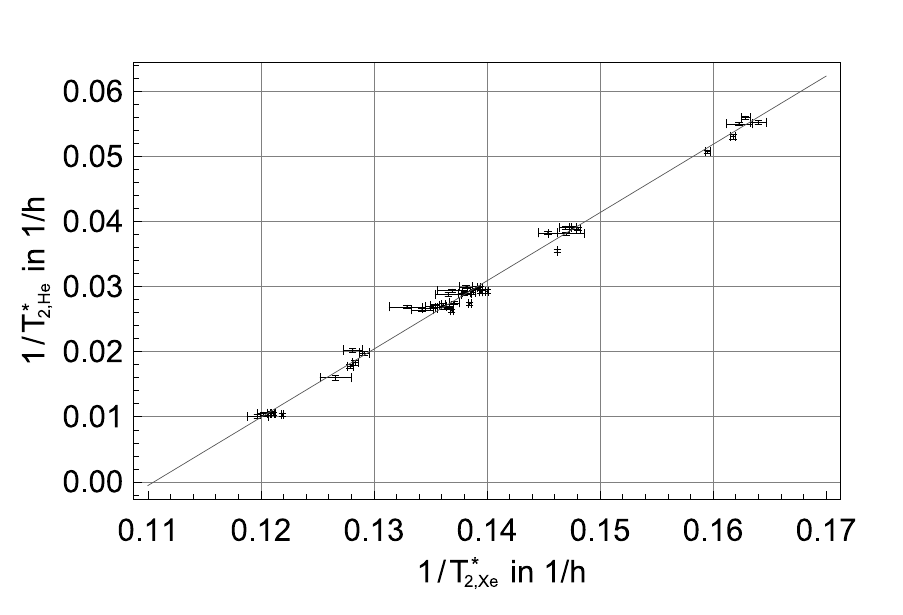}
\caption{Corresponding pairs of helium and xenon relaxation rates as magnetic field gradients are varied. Each point represents one pair of simultaneously measured helium and xenon relaxation rates. Solid line: straight line fit to the data according to Eq.~(\ref{eqn:T2*compare}).}
\label{fig:HevsXe}
\end{figure}
A straight line fit to the data finally gives
\begin{eqnarray}
m&=&1.049\pm 0.009
\label{eqn:fitresultm}
\end{eqnarray}
\begin{eqnarray}
k&=&(-0.116\pm 0.0013)~\text{h}^{-1}~~.
\label{eqn:fitresultk}
\end{eqnarray}
In order to confirm the validity of Eq.~(\ref{eqn:T2*}) one can compare the fit results with the corresponding values of $m$ and $k$  determined from independently measured quantities that enter on the right hand side of Eqs.~(\ref{eqn:k}) and (\ref{eqn:m}). For the ratio of the diffusion constants using the fit result (Eqs.~(\ref{eqn:fitresultm}) and (\ref{eqn:fitresultk})) and the precisely known ratio of the gyromagnetic ratios $\frac{\gamma_{He}}{\gamma_{Xe}}=2.75408159(20)$ \cite{Codata}, we get:
\begin{eqnarray}
\frac{D_{\text{He}}}{D_{\text{Xe}}}&=&\frac{\gamma^2_{He}}{m\gamma^2_{Xe}}=7.23\pm 0.07~~.
\label{eqn:Dratiomeasured}
\end{eqnarray}
In a gas mixture with $N_2$ as buffer gas, the resulting diffusion coefficients for  $^3$He and $^{129}$Xe are given by \cite{Mair}:
\begin{eqnarray}
\frac{1}{D_\text{He}}&=&\frac{p_\text{He}}{D^0_\text{He}}+\frac{p_\text{Xe}}{D^0_\text{He in Xe}}+\frac{p_{\text{N}_2}}{D^0_{\text{He in N}_2}}\\
\frac{1}{D_\text{Xe}}&=&\frac{p_\text{Xe}}{D^0_\text{Xe}}+\frac{p_\text{He}}{D^0_\text{Xe in He}}+\frac{p_{\text{N}_2}}{D^0_{\text{Xe in N}_2}}
\label{eqn:D}
\end{eqnarray}
with the diffusion coefficients of pure He and Xe: $D^0_\text{He}=(1.92 \pm 0.11)~\text{bar cm}^2\text{s}$ \cite{Barbe} and $D^0_\text{Xe}=(0.058 \pm 0.003)~\text{bar cm}^2\text{s}$ \cite{Acosta}, as well as the binary diffusion coefficients of the noble gas mixtures: $D^0_\text{He in Xe}=(0.610 \pm 0.031)~\text{bar cm}^2\text{s}$ \cite{Acosta}, $D^0_{\text{He in N}_2}=(0.771 \pm 0.039)~\text{bar cm}^2\text{s}$ \cite{Acosta}, $D^0_\text{Xe in He}=(0.548 \pm 0.023)~\text{bar cm}^2\text{s}$ \cite{Acosta,Hogervorst,Marrero,Malinauskas,Srivastava}, and $D^0_{\text{Xe in N}_2}=(0.128 \pm 0.004)~\text{bar cm}^2\text{s}$ \cite{Marrero,Trengove} (all values for $T=300$~K). Inserting these values into Eq.~(\ref{eqn:D}) results in the ratio of diffusion coefficients
\begin{eqnarray}
\frac{D_{\text{He}}}{D_{\text{Xe}}}&=&\frac{\left(24.03 \pm 0.96 \right)}{\left(3.55 \pm 0.10 \right)}=6.77\pm 0.33~~.
\label{eqn:Dratioliterature}
\end{eqnarray}
Within the error bars, this reproduces the fit result of Eq.~(\ref{eqn:Dratiomeasured}).\\
Furthermore, $T_{1,Xe}$ can be determined using the fit results for $k$ and $m$, and $T_{1,He}= (190\pm 10)$ h, which was also measured independently:
\begin{eqnarray}
T_{1,Xe}&=&\frac{m}{\frac{1}{T_{1,He}}-k}=(8.65 \pm 0.12)~\text{h}~~.
\label{eqn:T1Xe}
\end{eqnarray}
The longitudinal relaxation rate $1/T_{1,Xe}$ can be decomposed according to $1/T_{1,Xe}=1/T_{1,\text{wall}}+1/T_{1,\text{vdW}}$. These terms in turn can be determined from independent measurements published in \cite{Repetto1,Repetto2}, where the same sample cell was used. For the wall relaxation times of $^{129}$Xe, values between $17~\text{h} <T_{1\text{,wall}} <20~\text{h}$ were found which together with the expected relaxation time via Xe-Xe van der Waals (vdW) dimers, 
\begin{eqnarray}
\nonumber
T_{1,\text{vdW}}&=&T_1^{\text{Xe-Xe}} \left(1+r_{\text{N}_2} \frac{p_{\text{N}_2}}{p_\text{Xe}}\right)=(15.2 \pm 1.2)~\text{h}~~,\\
\label{eqn:T1vdW}
\end{eqnarray}
confirm the result of Eq.~(\ref{eqn:T1Xe}). The overall excellent agreement not only constitutes the direct experimental verification of Eq.~(\ref{eqn:T2*}), it also paves the way to use the free spin precession technique in order to determine very accurately magnetic field gradients from the measured $T_2^*$:\\
From the given diffusion coefficients of both noble gases in Eq.~(\ref{eqn:Dratioliterature}), we derive the respective numbers for the parameter $\lambda$ from Eq.~(\ref{eqn:lambda}), i.e., $\lambda_{He}\approx1.7 \cdot 10^{-4}$ and $\lambda_{Xe}\approx2.8 \cdot 10^{-5}$. For those values of $\lambda$ the corresponding pre-factor $a(\lambda)$ is essentially zero (see Fig. \ref{fig:ax}). So Eq.~(\ref{eqn:T2*}) is reduced to 
\begin{eqnarray}
\frac{1}{T_2^*}&=&
\frac{1}{T_1}+\frac{8R^4\gamma^2}{175D}|\vect{\nabla}B_z|^2
\label{eqn:T2*2}
\end{eqnarray}
from which the respective modulus of the field gradient value $|\vect{\nabla}B_z|$ can be derived. In case of $^3$He we find for $T_2^*=100$ h, which is about the value of the maximized  transverse relaxation time (see Fig.2): 
\begin{eqnarray}
|\vect{\nabla}B_z|=\left(5.6 \pm 0.4\right)~\text{pT/cm} ~~.
\end{eqnarray}
The uncertainty is essentially determined by the uncertainties of $D_{He}$ and the longitudinal relaxation time $T_1$.\\
In Fig.~\ref{fig:T2alpha} the extracted field gradient values from the measured transverse relaxation times $T_2^*$ of $^3$He are displayed, too. These values vary in the range $5.6~\text{pT/cm}~< |\vect{\nabla}B_z| < 17~\text{pT/cm}$ as the guiding magnetic field is rotated in the horizontal plane with respect to the residual field of BMSR-2.
The sensitivity to field gradient changes (monitoring) is solely determined by the accuracy with which $T_2^*$ can be measured. In the example given, $\delta|\vect{\nabla}B_z|\approx30$~fT/cm.
For higher magnetic field gradients, $T_2^*$ reduces accordingly. Given the same SNR, the data acquisition time to reach the same precision in $T_2^*$ determination is proportional to $\left(T_2^*\right)^{2/3}$. Thus, the response time to field gradient changes is greatly reduced.\\
\minisec{Discussion and Outlook}
We have demonstrated that hyperpolarized noble gas magnetometers based on the detection of free spin precession can be simultaneously used to detect magnetic field gradients with an accuracy in the sub pT/cm range. The observable is the transverse relaxation rate deduced from the exponential decay of the signal amplitude which directly depends on the square of absolute field gradient values across the spherical spin sample.  Due to the experimental setup and the performance of the measurements, we only had access to the longitudinal components $|\vect{\nabla}B_z|$ of the field gradient tensor. By specific settings of the weighting factor $a(\lambda)$ in Eq.~(\ref{eqn:T2*}), e.g., by changing the gas pressure, access to the transverse field gradient components is given with similar precision. Using pure $^3$He gas at $p_{He}= 1$ mbar, e. g., the value of the pre-factor $a(\lambda)$ approaches 0.5 for the same size of the sample cell.\\
The method can be further refined by using appropriate field gradient coils around the position of the spherical spin sample. From the known coil geometries and applied coil currents, well defined magnetic field gradients can be added to the residual and unknown field gradients at the sample position. Five linear independent sets of gradient coils are sufficient to determine the full tensor via $T_2^*$-measurements. In a forthcoming paper this measure to determine the traceless, symmetric second-rank field gradient tensor $\vect{G}$ will be discussed.\\
\minisec{Acknowledgements}
We want to thank our glass blower, Rainer Jera, for his high quality work in producing and repairing many cells and parts. This work was supported by Deutsche Forschungsgemeinschaft (DFG) under Contracts No. HE 2308/16-1 and SCHM 2708/3-1, by PRISMA cluster of excellence at Mainz, by Carl-Zeiss-Stiftung and by the Dutch Stichting FOM under programme 125 (Broken Mirrors and Drifting Constants).

\end{document}